# Spin Hall magnetoresistance effect from a disordered interface

*Sara Catalano[1*], Juan M. Gomez-Perez[1], M. Xochitl Aguilar-Pujol[1], Andrey Chuvilin[1,2], Marco Gobbi[1,2,3], Luis E. Hueso[1,2], and Fèlix Casanova[1,2]*

[1]CIC nanoGUNE, 20018 Donostia-San Sebastián, Basque Country, Spain

[2]IKERBASQUE, Basque Foundation for Science, 48009 Bilbao, Basque Country, Spain

[3]Centro de Física de Materiales CFM-MPC (CSIC-UPV/EHU), 20018 Donostia-San Sebastian, Basque Country, Spain

**ABSTRACT**

The Spin Hall magnetoresistance (SMR) emerged as a reference tool to investigate the magnetic properties of materials with an all-electrical set-up. Its sensitivity to the magnetization of thin films and surfaces may turn it into a valuable technique to characterize Van der Waals magnetic materials, which support long range magnetic order in atomically thin layers. However, realistic surfaces can be affected by defects and disorder, which may result in unexpected artifacts in the SMR, rather than the sole appearance of electrical noise. Here, we study the SMR response of heterostructures combining a platinum (Pt) thin film with the Van der Waals antiferromagnet MnPSe$_3$ and observe a robust SMR-like signal, which turns out to originate from the presence of strong interfacial disorder in the system. We use transmission electron microscopy (TEM) to characterize the interface between MnPSe$_3$ and Pt, revealing the formation of a few-nanometer-thick platinum-chalcogen amorphous layer. The analysis of the transport and TEM measurements suggests that the signal arises from a disordered magnetic system formed at the Pt/MnPSe$_3$ interface, washing out the interaction between the spins of the Pt electrons and the MnPSe$_3$ magnetic lattice. Our results show that damaged interfaces can yield an important contribution to SMR, questioning a widespread assumption on the role of disorder in such measurements.

**INTRODUCTION**

The spin Hall magnetoresistance (SMR) allows the surface magnetization of materials to be probed by measuring the magneto-transport response of a heavy metal (HM) interfaced with the magnetic compound.

The SMR effect relies on the spin-orbit interaction in the HM layer, coupling the spin and charge current by the spin Hall effect (SHE) and its reciprocal (inverse SHE). As a spin current is generated in the HM by the SHE, the spins reaching out the interface will be absorbed or reflected depending on the orientation of the magnetization ($M$) in the magnetic layer. Consequently, a corresponding charge current modulation, induced by inverse SHE, can be measured. More in detail, the spin current modulation varies as the vectorial product $\boldsymbol{m} \times (\boldsymbol{m} \times \boldsymbol{\mu_s})$, where $\boldsymbol{\mu_s}$ is the spin accumulation vector in the HM and $\boldsymbol{m} = (m_x, m_y, m_z)$ is a unit vector along $\boldsymbol{M}$ [1,2]. Thus, the longitudinal ($\rho_L$) and transverse ($\rho_T$) resistivity of the HM vary according to the equations:

$$\rho_L = \rho + \Delta\rho_0 + \Delta\rho_1(1 - m_y^2), \tag{1}$$

$$\rho_T = \Delta\rho_1 m_x m_y + \Delta\rho_2 m_z, \tag{2}$$

where $\rho$ is the Pt resistivity, $\Delta\rho_0$ accounts for a resistivity correction due to the SHE, and $\Delta\rho_1$ is the SMR amplitude, which depends on the real part of the spin-mixing conductance ($G_r$). The anomalous-Hall-like amplitude $\Delta\rho_2$ in Eq. 2 appears when $\boldsymbol{M}$ is out-of-plane, since $\boldsymbol{\mu_s}$ precesses around $m_z$, an effect quantified by the imaginary part of the spin-mixing conductance ($G_i$). Overall, the spin-mixing conductance $G_{\uparrow\downarrow} = G_r + iG_i$ determines the efficiency of the spin transport across the HM/magnet interface. Lately, it has been shown that an equivalent SMR effect also results from the interaction of a spin-orbit metal with an antiferromagnetic lattice, as the spin current at the interface is sensitive to the orientation of each probed magnetic sublattice[3–6]. The SMR equations can be extended to such a case by replacing ***m*** with the Néel vector ***n***, as described in more recent theoretical and experimental works[3,4,7,8]. As compared to other magnetoresistance (MR) effects, such as anisotropic MR (AMR) or the ordinary MR (OMR), SMR distinguishes for its symmetry, defined by equation (1) and (2).

Since its first observation[1], SMR has emerged as a powerful tool to characterize magnetic systems, especially thin films and surfaces, in alternative to volume sensitive magnetometry methods (SQUID, VSM)[9,10]. Notably, SMR may extend the capability of characterizing the magnetic behavior of exfoliated Van der Waals compounds, captivating the interest of the 2D materials community. The SMR effect has been observed in a system consisting of the van der Waals compound $MoS_2$ doped with Co impurities and Ta[11]. However, experimental studies of the SMR effect in proper layered magnets are lacking so far, while a variety of works on magneto-transport behavior in van der Waals magnetic compounds interfaced with heavy metals (Ta, Pt, W) have been reported, highlighting the importance of such heterostructures for spintronic applications[12–17]. We note here that, to date, most of such interfaces are assembled by sputtering of a heavy metal layer on top of the selected van der Waals material, even though their structural characterization is scarcely reported.

Naturally, to provide a proper material characterization, SMR measurements rely on the fabrication of a clean interface, preserving the magnetic properties of the studied compound[18–20]. In this context, the SMR amplitude has been associated with the achievement of high-quality interfaces[5,20], as it is commonly assumed that interfacial disorder, if extending over nanometer thickness, will wash out the spin signal. In this work, we show that such an assumption can be misleading and that a robust and large SMR-like magnetoresistance can be observed in HM/magnetic insulator bilayers affected by nanometer-thick interfacial disorder. We show that, in such a system, an anomalous increase of the SMR amplitude with the temperature can be observed, which can be explained reasonably by considering the transport behavior and magnetic properties of the interfacial layer. We note that similar anomalous temperature dependences of the SMR signal have been reported already but lacking a satisfactory explanation so far [5,21]. Moreover, we find that the interfacial disorder is directly caused by the sputtering deposition of the HM [22–24], which is commonly employed in the fabrication of HM/van der Waals interfaces. Notably, our work suggests that magneto-transport experiments on HM/van der Waals interfaces prepared with similar methods should be examined with caution.

## MATERIALS AND FABRICATION METHODS

For our experiment, we use platinum (Pt) as the HM layer, due to its large spin Hall angle and well characterized properties[25–28]. The magnetic counterpart consists of the Van der Waals compound $MnPSe_3$, a semiconductor with a relatively large bandgap [29], which exhibits antiferromagnetic order in-plane below a Néel temperature $T_N$ = 74±2 K [30]. More in detail, in the magnetic ground state, the $Mn^{2+}$ ions carry a magnetic moment S = 5/2 with intralayer antiferromagnetic exchange coupling with all nearest neighbors in the lattice [30,31]. The large bandgap ensures insulating transport properties of the compound over the full temperature range studied in this work (300 K – 10 K). Magnetization measurements on $MnPSe_3$ bulk crystal taken with a Vibrating Sample Magnetometer (VSM) confirm antiferromagnetic ordering in-plane below the expected $T_N$, as illustrated in *fig. 1a*. The crystalline quality of the exfoliated flakes is verified by Raman spectroscopy measurements, attesting the typical phonon fingerprint of MnPSe$_3$[31] (***fig. 1b***).

To fabricate the Pt/MnPSe$_3$ devices, MnPSe$_3$ flakes were mechanically exfoliated from a commercial crystal (supplied by HQ graphene) on a Si/SiO$_2$ substrate. The exfoliation was carried out in a glovebox inert environment. Afterwards, the exfoliated MnPSe$_3$ flakes, were rapidly transferred in an ultra-high vacuum (UHV) magnetron sputtering chamber ($P_{base}$=1 × 10$^{-8}$ mbar) and 5 nm of Pt was deposited on top of the flakes. The Pt film was then shaped into a Hall bar (channel length $L$ = 10um, width $w$ = 4 um) with standard electron-beam lithography followed by Ar-ion milling. Finally, the outer area of the Pt contacts was coated with a 100-nm-thick gold layer after a second lithography, to ensure continuity of the electrodes at the edge of the flake. An exemplifying image of one of the studied devices, consisting of a Pt Hall bar defined on top of a 100-nm-thick flake of MnPSe$_3$, is presented in ***fig. 1c***.

Magneto-transport measurements in Pt were carried out in a liquid-He cryostat at temperatures $T$ between 10 K and 300 K, externally applied magnetic fields $H$ up to 9 T, and a 360º sample rotation. The temperature dependence of the resistivity of the device is shown in ***fig. 1d***, confirming the expected transport characteristics of the Pt layer[25].

The quality of the Pt/MnPSe$_3$ interface was probed by (scanning) transmission electron microscopy (S)TEM, performed on a TitanG2 60–300 electron microscope (FEI Co., The Netherlands). The composition profiles were acquired in STEM mode utilizing energy dispersive X-ray spectroscopy (EDX) signal.

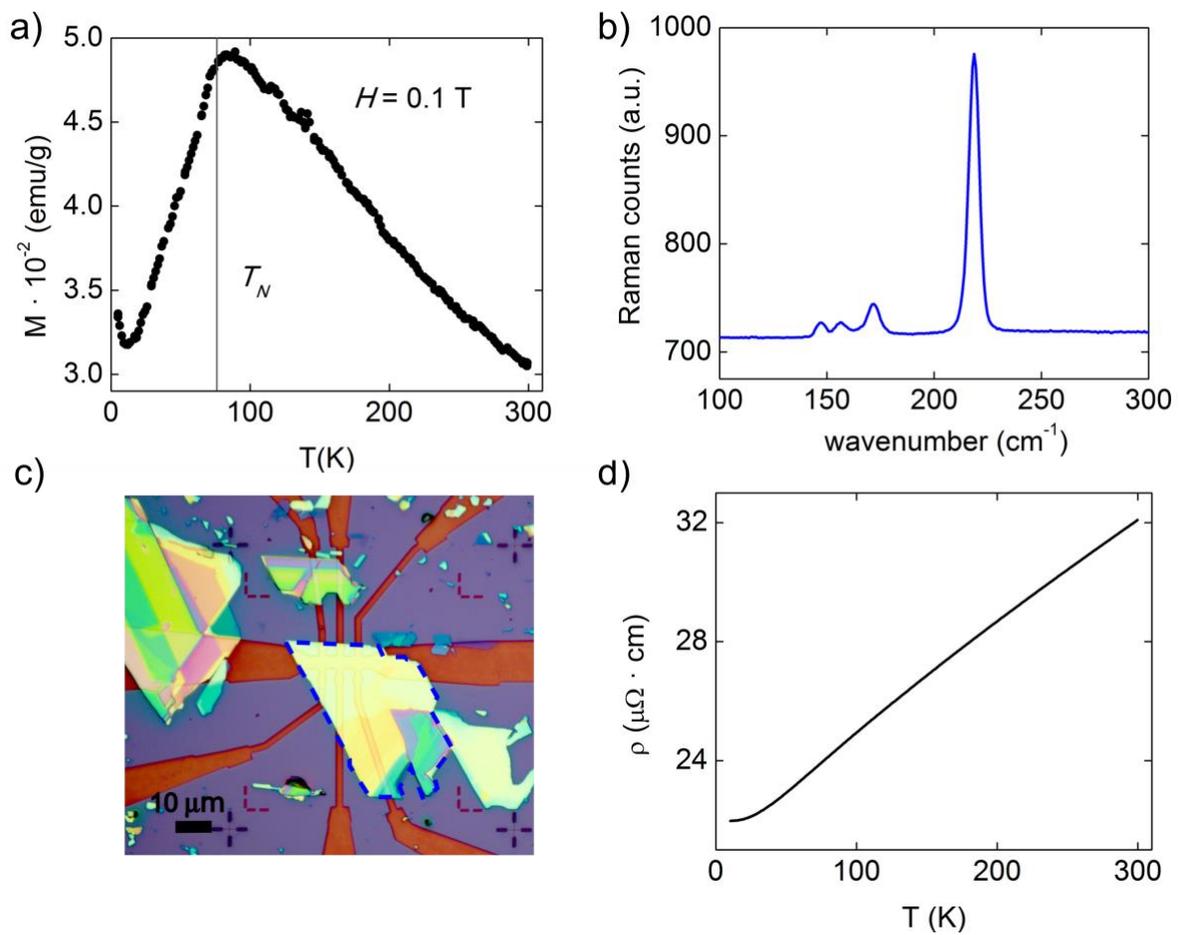

**Figure 1** (a) Magnetization of a bulk crystal of MnPSe$_3$ as a function of temperature at 0.1 T, measured by VSM; (b) Raman spectroscopy of an exfoliated MnPSe$_3$ flake; (c) Optical image of the device. The MnPSe$_3$ flake is outlined with the dashed blue line. Note that the 5-nm-thick Pt layer loses contrast on top of the exfoliated material; (d) Resistivity of the device shown in panel (c) as a function of temperature.
Figure 1

**RESULTS AND DISCUSSION**

We sketch, in the right panel of *fig. 2*, the measurements configuration. The device resistance was measured both in longitudinal ($R_L$) or transverse configuration ($R_T$). According to a common SMR convention, the *x*-axis is chosen along the current channel, the *y*-axis is parallel to the transverse arms of the Hall bar, and the *z*-axis is oriented along the out-of-plane direction, as illustrated. With such a choice of coordinates, a charge current flowing along *x* will generate a *y*-polarized spin accumulation at the Pt/MnPSe$_3$ interface. Then, $R_L$ and $R_T$ are recorded in two different modes: we will refer to the measurements performed as a function of the angle along the *xy* ($\alpha$), *yz* ($\beta$) and *xz* ($\gamma$) plane with a fixed external magnetic field ***H*** as angular-dependent magnetoresistance (ADMR) measurements and, correspondingly, to the measurements as a function of ***H*** at a fixed angle as field-dependent magnetoresistance (FDMR) measurements. The results of the ADMR measurements at $H = 9$ T, plotted as $\Delta \rho_L/\rho = [R_L(\alpha,\beta,\gamma) - R_L(90º)]/R_L(90º)$, are presented in *fig. 2*, for a temperature below $T_N$ of MnPSe$_3$ (50 K, *fig. 2a,b,c*) and well above $T_N$ (150 K, *fig. 2d,e,f*). For simplicity, we will discuss only the longitudinal measurements, the transverse ones being consistent with our findings (see Supporting Information). The ADMR shows a $\cos^2(\theta)$ dependence ($\theta = \alpha, \beta, \gamma$). In both the $\alpha$ and $\beta$ plane, for both T = 50 K and T = 150 K, the ADMR amplitude is on the order of $10^{-4}$, comparable with typical SMR amplitudes in Pt/YIG interfaces[18,20,32–34]. A much weaker MR modulation with a $\cos^2$ dependence is also observed in the $\gamma$ plane. At a first glance, we note that the amplitude of the ADMR exhibits the symmetry of the SMR, with a larger modulation as the field is swept from perpendicular ($\alpha, \beta = 0º$) to parallel ($\alpha, \beta = 90º$) to the spin accumulation $\boldsymbol{\mu_s}$ in the Pt layer. Instead, the weaker ADMR observed in the $\gamma$-plane can be accounted by the presence of an OMR contribution, which has already been observed in sputtered Pt films[35], [32] or to the presence of magnetic impurities in the Pt film, introducing a small but finite AMR effect. However, we stress here that we observe a dominant modulation in the α and β planes, consistent with the SMR symmetry constraints, which is at least one order of magnitude larger than the γ-plane modulation, and, as we will see, also presents a different temperature dependence.

In addition, we also prepared a control sample, replacing the heavy metal Pt film with sputtered copper (Cu), which is characterized by a very small spin-orbit coupling. In this case, we do not observe any SMR signal, confirming that the presence of a heavy metal with large spin Hall angle, such as Pt, is crucial for explaining the origin of the ADMR that we measure in the Pt/MnPSe$_3$ stack (see Supporting Information for a detailed description of the control experiment).

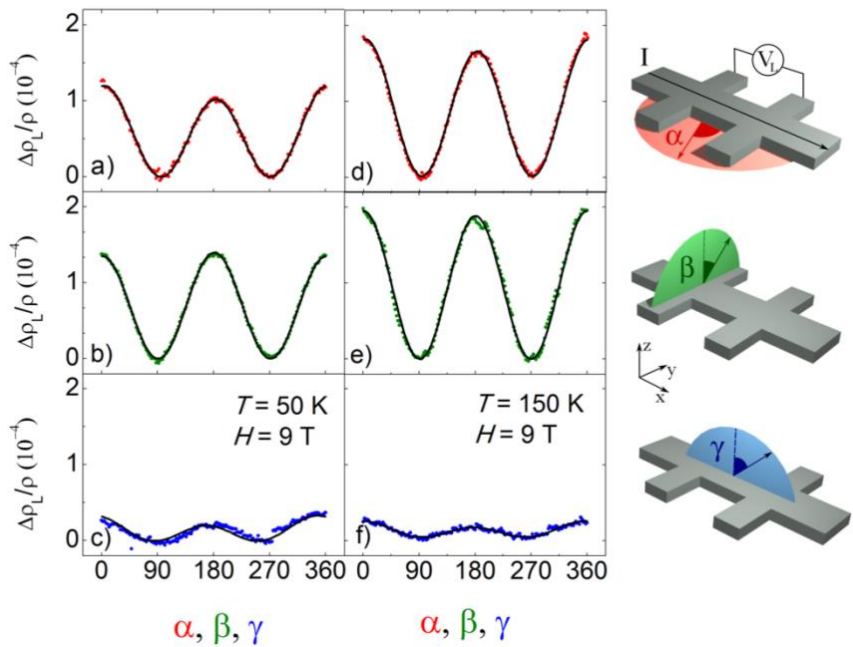

**Figure 2** ADMR measured at $H$ = 9 T in the α (red), β (green) and γ (blue) planes. (a,b,c) ADMR in the α, β and γ plane, respectively measured at T = 50 K. (d,e,f) ADMR measured at T = 150 K. The measurement set-up is sketched in the right panels, showing the planes wherein the external

field $H$ is rotated. The red, green, and blue semicircles correspond, respectively, to the $xy$ ($\alpha$), $yz$ ($\beta$) and $xz$ ($\gamma$) planes. The resistivity is measured in longitudinal configuration, by measuring $V_L$ while a current is applied parallel to the x-axis, as sketched in the top right panel.

At first, it would be tempting to assign the large ADMR amplitude measured in the $\alpha$ and $\beta$ planes to the interaction between $\boldsymbol{\mu_s}$ in Pt and $\boldsymbol{n}$ describing the antiferromagnetic ordering of MnPSe$_3$. Yet, further measurements reveal a more complex picture.

In figure 2, we show that the observed SMR-like signal is present both below (***fig. 2a,b,c***, T = 50 K) and above (***fig. 2d,e,f***, T = 150 K) $T_N$. More in detail, the ADMR amplitude displays an unusual temperature dependence, as it increases with the temperature, reaching a saturation value of ~2 × 10$^{-4}$, as illustrated in ***fig. 3a***.

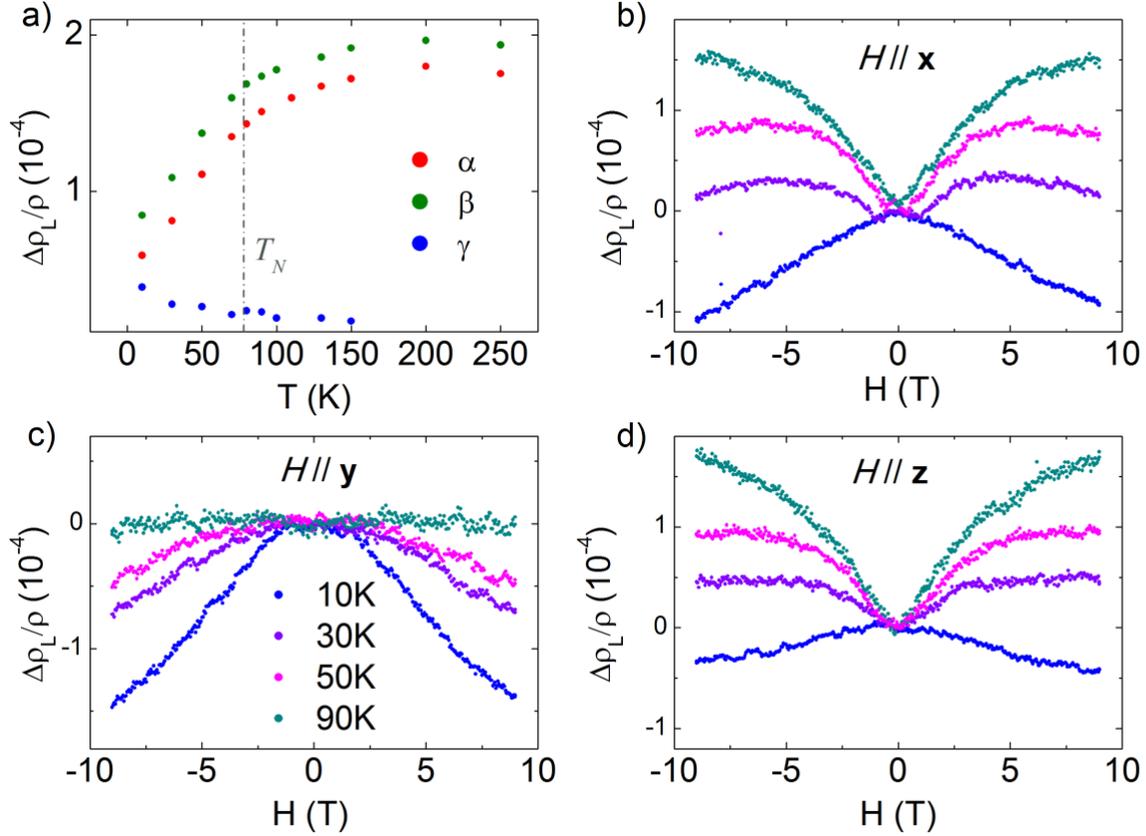

**Figure 3:** (a) Temperature dependence of the SMR amplitude obtained from the ADMR measured in the α (red circles), β (green circles), and γ (blue circles) plane at 9 T. The vertical dashed line corresponds to the Néel temperature of MnPSe$_3$. (b, c, d) FDMR measured, respectively, with *H*//x (b), *H*//y (c), and *H*//z (d) at different temperatures, as indicated in the color-legend.

Moreover, FDMR measurements reveal that the signal amplitude does not saturate for fields up to 9 T. As illustrated in ***fig. 3c***, when *H* is applied along *y*, i.e. parallel to $\boldsymbol{\mu_s}$, we observe a clear negative MR at low temperature (T = 10 K), which gradually flattens as the temperature is increased. When *H* is applied along *x* (*fig.3b*) and *z* (*fig.3d*), i.e. orthogonal to $\boldsymbol{\mu_s}$, three MR regimes are observed. At the lowest measured temperature (T = 10 K), the MR is negative at all measured fields. As the temperature increases (T ≥ 30 K), the MR signal first increases rapidly with the field and eventually decreases again for $H \geq 5$ T. Finally, for higher temperatures (T $\geq$ 90 K), the MR keeps increasing up to the highest measured field.

From such observations, we can already conclude that the origin of the SMR-like signal cannot lie in the antiferromagnetic structure of MnPSe$_3$. In fact, if the modulation was related to the emergence of an antiferromagnetic phase, we would expect a decrease of the SMR amplitude above $T_N$ as MnPSe$_3$ turns into a paramagnet, instead of a monotonic increase[4]. While short range spin-spin correlations can persist even above $T_N$ in layered magnetic compounds[36,37] and may be thought to contribute to the SMR amplitude, we see that the signal persists at least up to 250 K, a temperature at which such correlations cannot play a role. In addition, the different regimes observed in the FDMR suggest the presence of competing magnetic interactions, which cannot be explained by the SMR theory considering only the emergence of antiferromagnetic order in the system below $T_N$ [2]. In fact, the FDMR signal expected for the magnetic ground state of MnPSe$_3$ should exhibit a monotonic increase (decrease) as the external magnetic field is applied along the $x$ and $z$ ($y$) axes.

Alternatively, one may invoke a Hanle MR (HMR) contribution to the measurements to explain the presence of a SMR-like signal above $T_N$. Indeed, HMR shares the same symmetry of SMR and, to the leading order corrections, increases as $(H/D)^2$, where $D$ is the electron diffusion constant[33,38]. Yet, the amplitude of the measured signal is at least one order of magnitude higher than the HMR amplitude observed at 9 T in Pt films deposited on top of paramagnetic systems[33] and, in addition, the field dependence deviates from a parabolic law. Thus, we rule out the HMR as the origin of the observed signal (see Supporting Information for a more detailed analysis).

To further elucidate the origin of the complex magneto-transport data, we used TEM to characterize the interface of the system. *Fig. 4a* shows a TEM image of a cross-section of the interface region in a pristine Pt/MnPSe$_3$ bilayer, obtained right after the deposition by sputtering of the 5-nm-thick Pt film on top of as-exfoliated MnPSe$_3$. TEM data of a device studied by transport measurements is presented in the Supporting Information, reproducing completely equivalent results. *Fig. 4a* reveals the crystalline MnPSe$_3$ flake at the bottom of the stack, a 2-nm-thick amorphous layer, and the Pt polycrystalline film on top. EDX in STEM mode (signals for Pt, Mn and Se are overlaid in the image) shows that the interfacial amorphous layer consists of Pt (green solid line) intermixed with Se (blue solid line), whereas the Mn cations (red solid line) from MnPSe$_3$ do not exhibit interdiffusion. Thus, the analysis of the TEM cross-section evidences that the Pt/MnPSe$_3$ contact differs strongly from an ideal tight atomically flat interface. As the unit cell of the van der Waals tri-phosphate consists of a hexagonal sublattice of Mn cations sandwiched between two Se-chalcogen layers, it is likely that a part of the Se ions of the top layer segregate into the Pt film, resulting in the formation of Se-vacancies at the surface. The interdiffusion of the Se atoms into the Pt film results in an amorphous interfacial layer, interposed between the MnPSe$_3$ surface and the crystalline Pt film. We note here that Pt and Se can form two stable stoichiometric compounds, namely, Pt$_5$Se$_4$ and PtSe$_2$, yet the TEM image does not show the formation of crystallites or any signature of an ordered phase. Moreover, the equilibrium phase diagram for Pt-Se show poor solubility, so that it is unlikely that a Pt-Se alloy is formed[39].

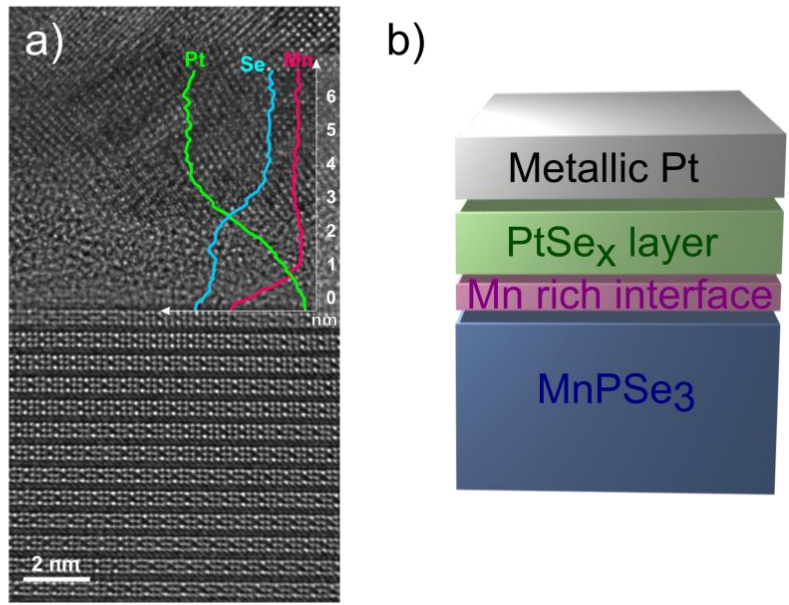

**Figure 4:** (a) High resolution TEM image of the Pt/MnPSe$_3$ bilayer cross-section, revealing a ~2-nm-thick granular layer interposing between the Van der Waals compound and the Pt film. The inset shows the corresponding chemical profile as revealed by EDX analysis. (b) Sketch of the proposed model of the heterostructure composition. A thin magnetic layer (magenta) and a semiconducting granular layer (green) interpose between the metallic Pt film on top and the bulk crystalline MnPSe$_3$ at the bottom.

The TEM characterization highlights a counterintuitive effect: we observe a relatively strong SMR signal, comparable to the value reported for clean Pt/YIG interfaces[1,20,40] in a system affected by interfacial disorder, extending for over 2 nm. Considering the presence of such an amorphous layer, we discuss here the possible mechanism(s) consistent with our observations, in particular accounting for (i) the temperature dependence of the ADMR amplitude, (ii) the origin of the SMR signals, and (iii) the evolution of the FDMR curves. We note here that the development of a proper model describing such a system is far from trivial and would require a deeper investigation of the interface structural and chemical details.

The EDX profile shows that, instead of an ideal Pt/MnPSe$_3$ bilayer, our sample consists of a thick MnPSe$_3$ layer, an intermediate layer composed of Pt grains and Se impurities, which we label as PtSe$_x$ layer, and a top Pt film. While the exact composition of the intermixed layer is unknown, we can formulate two reasonable assumptions: (i) the interface between the bulk MnPSe$_3$ layer and the PtSe$_x$ layer is characterized by the presence of uncompensated magnetic moments, most likely coming from the Mn atoms; (ii) the PtSe$_x$ layer exhibits semiconducting transport behavior, which is a reasonable assumption for nanometer thick amorphous films[41–43]. We provide an exemplifying sketch of our hypothesis in *fig. 4b*. We speculate that such a picture explains the temperature dependence of the ADMR amplitude. The SMR signal originates from the magnetic moments present at the interface between the PtSe$_x$ interlayer and bulk MnPSe$_3$. The semiconducting character of the former permits a more efficient transport of the charge and spin currents generated in the upper Pt film at higher temperature. Therefore, the SMR amplitude decreases as the resistance of the amorphous layer increases, that is, as the temperature decreases, in agreement with the measured temperature dependence of the ADMR. Additionally, the amorphous layer itself may contribute to the total spin current, as it hosts Pt atoms. We note that analogous temperature dependence of SMR amplitudes have been already reported in literature for interfaces between Pt and antiferromagnetic oxides[5,21], even though a satisfying explanation for such a behavior has not been provided so far. Hereto, we suggest that the presence of a thin semiconducting PtSe$_x$ film, interposing between the Pt film and the magnetic layer, can account for the anomalous temperature dependence of the signal.

As for the microscopic origin of the SMR signals, the most plausible source of magnetic moments are the Mn atoms from MnPSe$_3$, which in the bulk compound carry 5 $\mu_B$ (S = 5/2). Alternatively, one can speculate that the PtSe$_x$ layer is itself magnetic, taking into account recent reports of room temperature ferromagnetism in Pt-vacancy rich PtSe$_2$[44,45]. In both cases, we suggest that our heterostructures can be described as illustrated in *fig. 4b*, as a stack comprising the bulk MnPSe$_3$ layer at the bottom, a thin magnetic system at the surface of MnPSe$_3$, a semiconducting PtSe$_x$ layer and the polycrystalline Pt film at the top.

Next, we highlight that the observed FDMR curves do not resemble the SMR expected from the interaction with a ferromagnetic or ferrimagnetic system, which should saturate together with the magnetization of such a system[1]. Instead, analogue non-saturating FDMR curves have been observed in Pt films interfaced with a low dimensional Heisenberg ferromagnet,[46] for which the short-range exchange interactions are much stronger than for a typical paramagnet, resulting in an enhancement of the spin-mixing conductance and, thus, of the SMR amplitude. Alternatively, the lack of saturation can also be attributed to the formation of a spero- or sperimagnetic system at the interface, as such magnetic phases typically occur in amorphous compounds[47–49]. In such a picture, the uncompensated Mn magnetic moments would be frozen into random orientations with respect to each other, determined by the local exchange and anisotropy energy terms, each moment exhibiting a finite time-average. The magnetization curve of spero- and sperimagnets does not exhibit saturation even for very large fields, depending on the strength of the system exchange and anisotropy interactions. Thus, we propose that the amorphous layer formed at the Pt/MnPSe$_3$ interface corresponds to one of the described magnetic systems.

Finally, we note that the FDMR curves display a peculiar evolution through different temperatures. For $H$ oriented along the $y$ axis, the MR is negative at low temperature and gradually flattens as the temperature is increased. In contrast, for $H$ oriented along either the $x$ or $z$ axis, the MR display a non-monotonic behavior, with the slope changing as the external field is swept above $\approx 2.5$ T and gradually increasing as the temperature is raised. Such a behavior suggests the presence of competing magneto-transport mechanisms in our system. Most likely, together with the SMR effect we may observe an additional AMR contribution arising from spins localized in the conducting layer, which is consistent with the observation of a small $\cos^2(\gamma)$ modulation of the MR (***fig. 2c,f***).

**CONCLUSIONS**

SMR measurements are exceptionally sensitive to interface details and provide a powerful transport-based method to characterize magnetic systems. Ideally, they could help in characterizing the magnetic phases hosted by van der Waals compounds. In this work, we study the SMR fingerprint of heterostructures combining a van der Waals magnet ($MnPSe_3$) with a Pt layer. We measure a robust SMR signal, which shows an unusual increase with the temperature. We demonstrate that such a SMR signal stems from an amorphous layer at the Pt/$MnPSe_3$ interface, which unexpectedly host a magnetically ordered phase. We show that the interface amorphization is a consequence of the damage induced by the deposition by sputtering of the Pt film on top of the van der Waals compound. We suggest that the temperature dependence of the signal can be ascribed to a semiconducting behavior of the amorphous interface. Notably, such a picture may adequately explain comparable temperature dependence of the SMR reported recently, independently of the details of the studied systems.

Considering the increasing interest for spintronic applications relying on HM/van der Waals magnetic compounds heterostructures, we believe that our results offer important insights.

The deposition of heavy metals such as Pt, Ta or W by sputtering is a widely employed fabrication method. Our measurements attest that such a method can be detrimental to the quality of the fabricated interfaces. Moreover, we show that such damaged interfaces can provide an unpredicted and meaningful contribution to the magneto-transport of HM/van der Waals heterostructures and therefore a proper structural characterization should be considered when studying them.

**Acknowledgements**

The authors would like to thank Vitaly N. Golovach for fruitful discussions. The authors also thank SGIker Medidas Magnet- icas Gipuzkoa (UPV/EHU/ ERDF, EU) for the technical and human support. This work is supported by the European Union H2020 under the Marie Sklodowska-Curie Actions (796817-ARTEMIS and 748971-SUPER2D) and by the Spanish MICINN under project nos. RTI2018-094861-B-I00, PID2019-108153GA-I00, and the Maria de Maeztu Units of Excellence Programme (MDM-2016-0618 and CEX2020- 001038-M). J.M.G.-P. and M.X.A.-P. thank the Spanish MICINN for a Ph.D. fellowship (grants no. BES-2016-077301 and PRE-2019-089833, respectively). M.G. acknowl- edges support from la Caixa Foundation for a Junior Leader fellowship (grant no. LCF/BQ/PI19/11690017).

## ASSOCIATED CONTENT

**Supporting Information**.

Supporting Information includes:

- ADMR and FDMR of control sample consisting of $TaO_x$/Ti/$CuMnPSe_3$; ADMR measured in transversal configuration in the $\alpha$ plane, at T = 50 K, 90 K, 150 K; FDMR measured in transversal configuration in the $\alpha$ plane at T = 50 K and T = 90 K; TEM cross section and EDX profile of a measured device; discussion of the HMR mechanism

## AUTHOR INFORMATION

**Corresponding Author**

*s.catalano@nanogune.eu

**Author Contributions**

The manuscript was written through contributions of all authors. All authors have given approval to the final version of the manuscript.

## ACKNOWLEDGMENT

This work is supported by the European Union H2020 under the Marie Sklodowska-Curie Actions (796817-ARTEMIS and 748971-SUPER2D) and by the Spanish MICINN under the Maria de Maeztu Units of Excellence Programme (MDM-2016-0618) and under project No. RTI2018-094861-B-100. J.M.G.-P. and M.X.A.-P. thank the Spanish MICINN for a Ph.D. fellowship (Grants No. BES-2016-077301 and PRE-2019-089833, respectively). M.G. acknowledges support from la Caixa Foundation for a Junior Leader fellowship (Grant No. LCF/BQ/PI19/11690017).

# Supporting Information for

# Spin Hall magnetoresistance effect from a disordered interface

Sara Catalano[1*], Juan M. Gomez-Perez[1]†, M. Xochitl Aguilar-Pujol[1], Andrey Chuvilin[1,2], Marco Gobbi[1,2,3], Luis E. Hueso[1,2], and Fèlix Casanova,[1,2]

[1]CIC nanoGUNE, 20018 Donostia-San Sebastián, Basque Country, Spain

[2]IKERBASQUE, Basque Foundation for Science, 48009 Bilbao, Basque Country, Spain

[3]Centro de Física de Materiales CFM-MPC (CSIC-UPV/EHU), 20018 Donostia-San Sebastian, Basque Country, Spain

KEYWORDS. Spintronics, 2D materials, Spin Hall Magnetoresistance, Interfaces

**Corresponding Author**

*s.catalano@nanogune.eu

**S1: Magneto-transport in control sample (fig. S1, S2 and S3)**

To verify the SMR origin of the MR signal measured in the Pt/MnPSe$_3$ heterostructures, we performed a control experiment where we replace the Pt layer with a Cu film, as Cu exhibit very low spin-orbit coupling. Also in this case, we use DC magnetron sputtering for the metal deposition. By adopting again sputtering deposition, we aim at fabricating a metal/van der Waals interface analogue to the studied Pt/MnPSe$_3$ device, that is exhibiting similar damage. Thus, by replacing Pt with Cu we can verify that only in the case of having a heavy metal such as Pt on top the SMR signal can be probed.

We found out that Cu does not grow continuously on the surface of the MnPSe$_3$ layer (*fig.S1*), so that we sputtered a 2-nm-thick Ti buffer layer on top of the surface of MnPSe$_3$ prior to the Cu deposition. The Cu/Ti/MnPSe$_3$ stack is then capped with 2 nm of insulating TaO$_x$ to prevent oxidation of the Cu film. The final heterostructure consists of TaO$_x$(2nm)/Cu(10nm)/Ti(2nm)/MnPSe$_3$. The thickness of Cu is chosen to be 10 nm, as we found that sputtered Cu films of such a thickness exhibited good conductivity properties. We note that the Cu film thickness is slightly higher than the thickness of Pt in the Pt/MnPSe$_3$ devices (5nm), however, due to the long spin diffusion length of Cu, we believe that such a change should not affect the results of the control measurements. To measure the magneto-transport properties of the stack, the heterostructure is shaped into a Hall bar (length = 6μm, width = 1 μm) by electron-beam lithography (EBL).

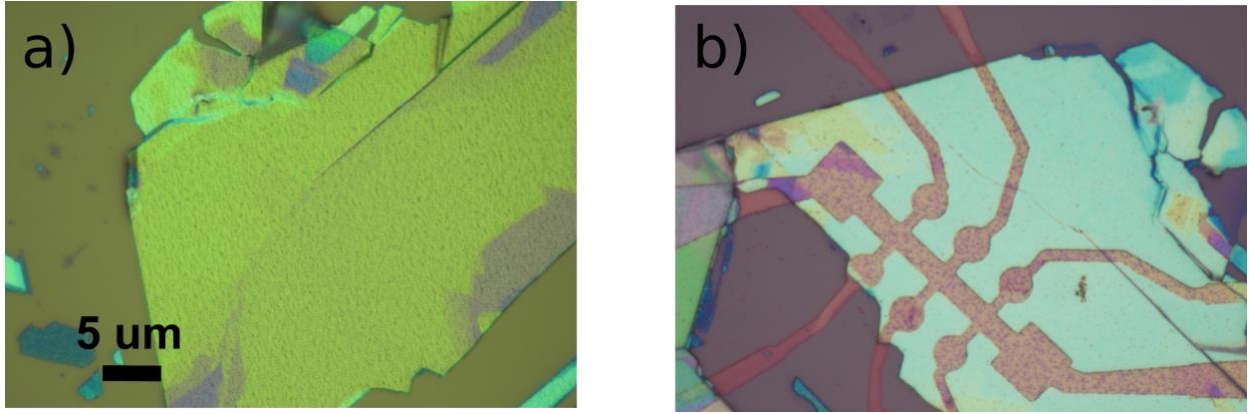

***Figure S1***. *Optical microscope images of MnPSe$_3$ flake after the deposition of 10 nm of Cu by DC magnetron sputtering. **(a)** Surface of MnPSe$_3$ after sputtering of Cu. The surface roughness is ascribed to the discontinuous growth of Cu. Note that Cu is sputtered everywhere, but only the surface of MnPSe3 looks rough. **(b)** Optical image of the final Cu Hall bar defined by EBL and ion-milling.*

In ***fig. S2***, we plot the temperature dependence of the resistivity of the sample, which shows metallic behavior. In ***fig. S3b*** we show the ADMR measured at 10 K, with an applied external field $H = 9$ T, in the α, β, and γ plane. We observe a clear angle dependent modulation in the β and γ plane, with identical amplitude. In contrast, the α plane measurements are dominated by resistance fluctuations. Since the amplitude of the modulation does not change from the beta to the gamma plane, we can immediately conclude that no SMR is present in the control sample, since the SMR effect vanishes in the γ plane. In ***fig. S3a*** we show the FDMR of the sample at 10K. We observe a sizeable MR as we measure the resistance as a function of an out of plane field ($H$ along the **z** axis), which is the origin of the observed ADMR, whereas the FDMR measured with the field $H$ parallel to the **x** axis matches with the FDMR measured with $H$ parallel to the **y** axis. Finally, we repeat the measurements at 100K. At this temperature, the ADMR exhibit a similar behavior with a smaller amplitude (***fig. S3c***). Such a temperature dependence of the signal also differs from the case of the Pt/MnPSe$_3$ interface, for which the signal amplitude increases with the temperature. We ascribe the observed negative magnetoresistance to the appearance of weak localization in such a system[50].

Thus, we conclude that the MR signal observed in the Pt/MnPSe$_3$ heterostructures can safely be ascribed to the SMR mechanism.

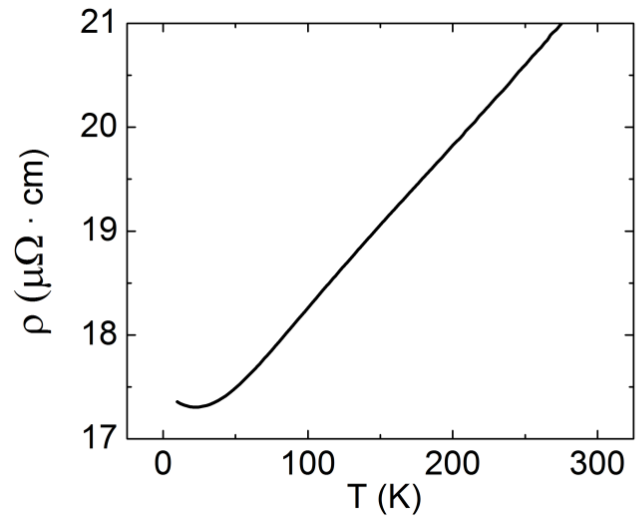

*Figure S2. Temperature dependence of the resistivity of the control sample TaO$_x$(2nm)/Cu(10nm)/Ti(2nm)/MnPSe$_3$.*

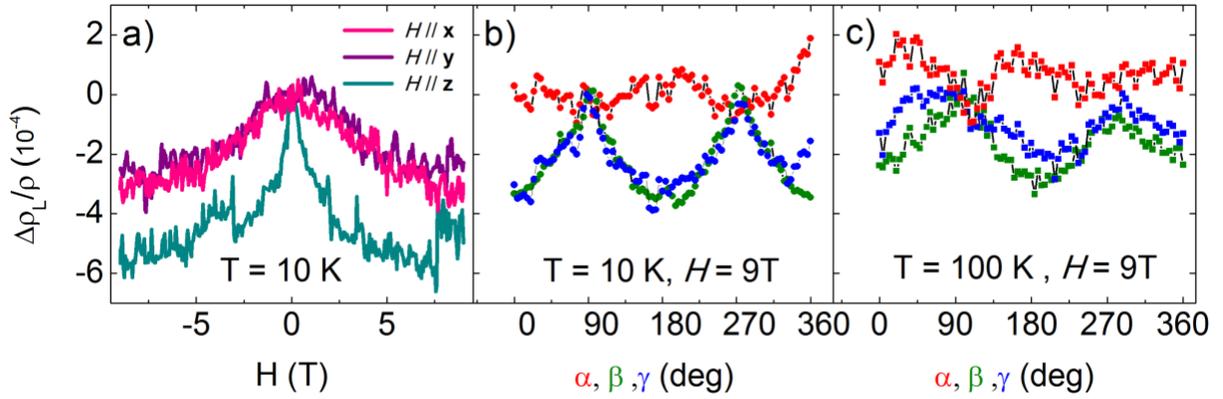

***Figure S3.*** *(a) FDMR of the control sample measured at T = 10 K, for H aligned along the **x** (magenta line), **y** (purple line), and **z** (cyan line) axes. (b) Angular dependence of the MR measured in the α, β, and γ planes at T = 10 K and H = 9 T. (c) Same as panel b), but at T = 100 K.*

**S2: Magneto-transport in transversal configuration (fig. S4 and fig. S5)**

In transverse configuration a current is passed through the Hall bar channel and the transverse voltage $V_T$ is measured across one Hall bar arm, that is, in the direction perpendicular to the current path, as sketched in ***fig. S4 d)***. Measurements in this configuration results in a clean signal when the magnetic field is swept in-plane (α-plane). Instead, measurements in the β– and γ-plane are affected by strong Hall effect contribution.

***Figs. S4a-c*** present the ADMR measured in this configuration at different temperatures (T = 50 K, 90 K and 150 K, respectively). The results agree with the longitudinal configuration measurements, showing the SMR amplitude increasing up to $\Delta\rho/\rho = 2 \times 10^{-4}$ (peak-to-peak amplitude) above the Néel Temperature ($T_N$ = 74 K) of MnPSe$_3$ (panel (a) T = 50 K, panel (b) T = 90 K and (c) T = 150 K).

*Fig.S5* shows the FDMR measured in transverse configuration at two different temperatures. In this case, the magnetic field is applied at fixed angles, α = 45° and α = 135°, which corresponds, respectively, to the maximum and minimum of the ADMR in transverse configuration.

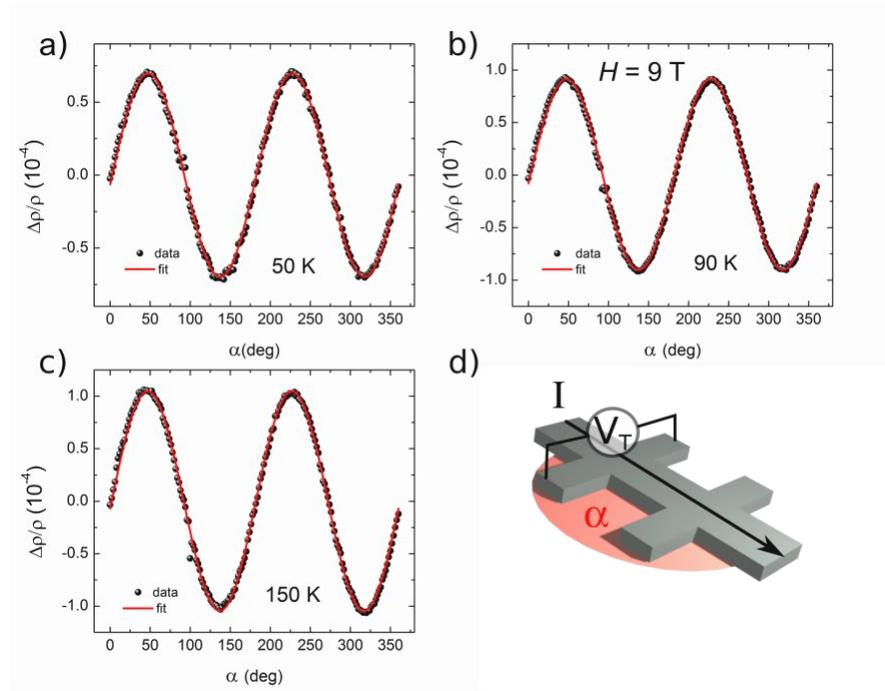

**Figure S4**. ADMR measured in transversal configuration in the α-plane, at H = 9 T, at T = 50 K *(a)*; T = 90 K *(b)*; T = 150 K *(c)*. The transversal configuration measurement scheme is illustrated in panel *(d)*.

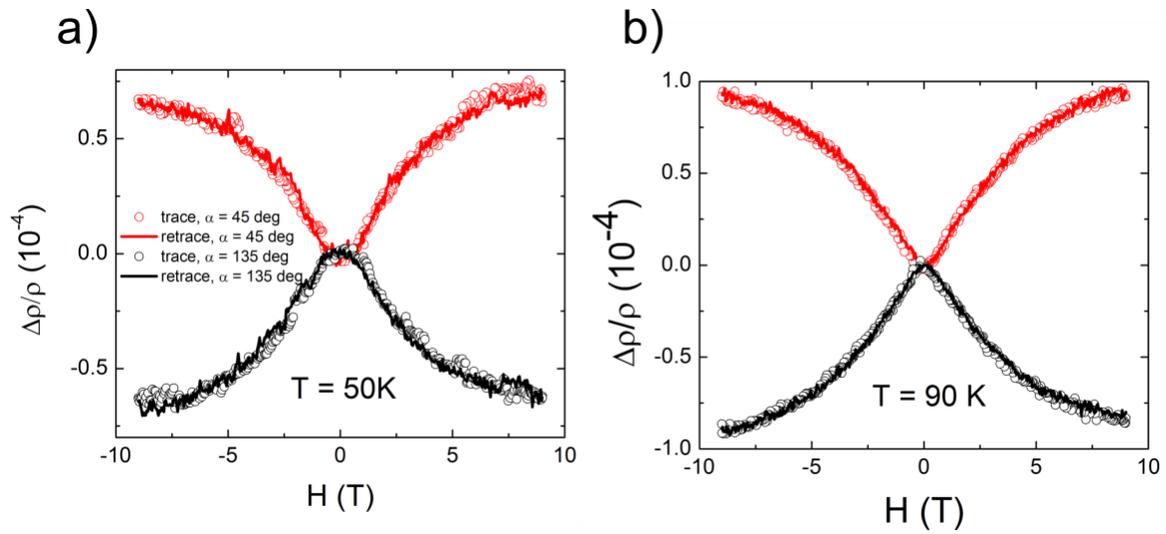

*Figure S5*. FDMR measured in transversal configuration at **(a)** $T = 50$ K and **(b)** $T = 90$ K. The SMR amplitude does not show saturation up to $H = 9$ T. The SMR-gap is absent both below and above the $T_N$ of MnPSe$_3$.

## S3: TEM cross section (fig. S6)

*Fig. S6* presents the TEM cross section of one of the measured devices. The results are equivalent to those presented in the main text.

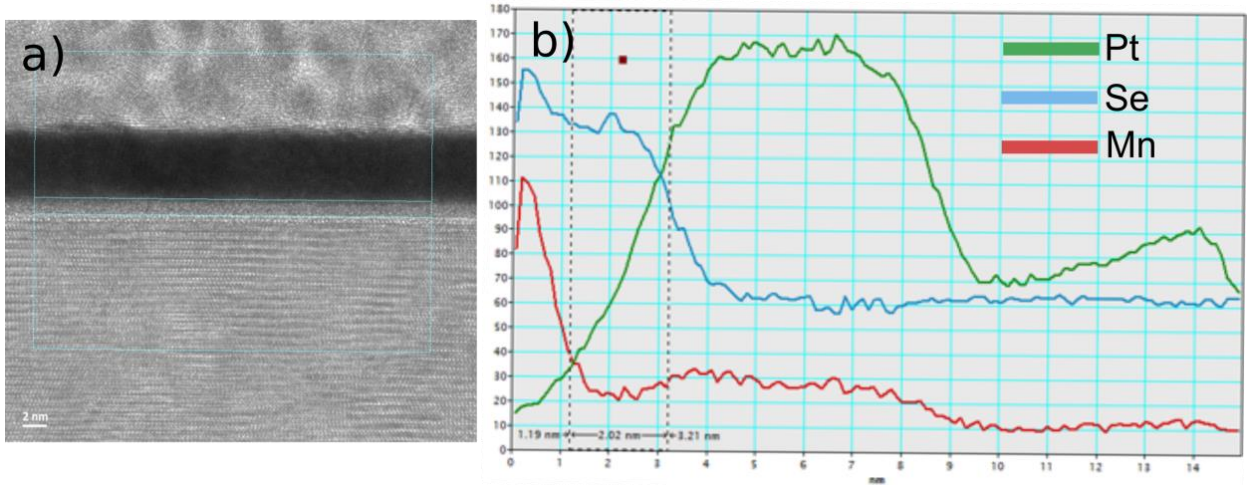

**Figure S6**. *(a) TEM cross section of one of the measured devices. (b) EDX profile. A granular layer of about 2 nm thickness is interposed between the MnPSe₃ crystal and the Pt layer sputtered on top (black contrast in panel a). The polycrystalline layer above the sputtered Pt consists of a second Pt film deposited on top of the sample to perform the TEM analysis. The chemical profile reveals that the granular layer is composed of Pt and Se atoms, consistent with the results shown in the main text.*

## S4: Hanle Magnetoresistance (HMR)

HMR shares the same symmetry of SMR and, to the leading order corrections, increases as $(H/D)^2$, where $D$ is the electron diffusion constant. More in detail, in sputtered Pt films, $D \approx 10^{-5}$ m$^2$s$^{-1}$, which satisfies the condition $\mu_0 H \ll \frac{\hbar}{g\mu_B \lambda^2} \cdot D$, where $\hbar$ is the reduced Planck constant, $g$ is the g-factor, $\mu_B$ is the Bohr magneton and $\lambda$ is the spin diffusion length of Pt. In the Pt/MnPSe$_3$ devices, the field dependence departs from a simple parabolic law, indicating that at least $\mu_0 H \gg \left(\frac{\hbar}{g\mu_B \lambda^2}\right) \cdot D$, which is not realistic in Pt sputtered films.

Thin Cu Films Deposited on Mica. *Scientific Reports 2021 11:1* **2021**, *11* (1), 1–10. https://doi.org/10.1038/s41598-021-97210-w.